\documentclass[12pt]{article}
\usepackage{graphics}
\usepackage{graphicx}
\DeclareGraphicsExtensions{.pdf}
\usepackage{float} %Include figure filesusepackage{graphicx} %Include figure files
\usepackage{amsmath}
\usepackage{slashed}
\usepackage{amsfonts}   
\usepackage{amssymb}

\begin{document}
\date{}
%%%%%%%%%%%%%%%%%%%%
\title{{\bf{\Large Matrix models and non-Abelian T dual of $AdS_5 \times S^5 $}}}
%%%%%%%%%%%%%%%%%%%%
\author{
 {\bf {\normalsize Dibakar Roychowdhury}$
$\thanks{E-mail:  dibakarphys@gmail.com, dibakar.roychowdhury@ph.iitr.ac.in}}\\
 {\normalsize  Department of Physics, Indian Institute of Technology Roorkee,}\\
  {\normalsize Roorkee 247667, Uttarakhand, India}
\\[0.3cm]
}
%\date{}

\maketitle
%%%%%%%%%%%%%%%%%%%%%%%%%%%%%%%%%%%%%%%%%%%%%%%%%%%%%%%%%%%%%%%%%%%%%%%%%%%%
\begin{abstract}
We compute Euclidean Wilson loops for BMN Plane Wave Matrix Models. The stringy counterpart of these Wilson loops corresponds to various semi-classical open string embedding that probe a class of half BPS geometries in Type IIA supergravity. These geometries fall under the general category of Lin-Lunin and Maldacena (LLM). As a special class of solutions within the LLM category, we construct Wilson operators corresponding to the non-Abelian T dual (NATD) of $ AdS_5 \times S^5 $. Our analysis shows close agreement between matrix model and string theory calculations.
\end{abstract}
%%%%%%%%%%%%%%%%%%%%%%%%%%%%%%%
\section{Overview and motivation}
Understanding the interplay between the open string solutions in $ AdS_5 \times S^5 $ and the dual Wilson loops \cite{Maldacena:1998im}-\cite{Ryang:2013yna} had witnessed a steady progress over the past few decades. All these results are promising and are therefore worth exploring in a different context of gauge/string correspondence. For example, investigating the counterpart of some of these Type IIB open string solutions in the context of Type IIA/ Plane Wave Matrix Model\footnote{It is considered to be the \emph{massive} deformation of the BFSS matrix model \cite{Banks:1996vh}. } (PWMM) correspondence might shed light beyond our current understanding of the Maldacena conjecture in a Type IIB set up \cite{Maldacena:1997re}. 

The purpose of the present paper is therefore to address some of these related issues (as alluded to the above) considering open string dynamics in Type IIA supergravity backgrounds preserving 8 (complex) supercharges \cite{Lin:2004nb}-\cite{Ling:2006up} and to find out an interpretation in terms of the dual BMN matrix models at strong coupling. The classification of these Type IIA geometries, as initiated due to Lin-Lunin and Maldacena (LLM) in \cite{Lin:2004nb}, has given rise to a plethora of QFTs preserving $ \mathcal{N}=2 $ SUSY. These supersymmetric QFTs naturally include the PWMM that was studied previously by authors in \cite{Berenstein:2002jq}-\cite{Dasgupta:2002hx}.

Among these classes of Lin--Lunin and Maldacena (LLM) Type IIA geometries, the non-Abelian T dual\footnote{The non-Abelian T duality (NATD) which (was introduced as a \emph{tree} level symmetry of the bosonic sigma models) essentially corresponds to a set of transformations between the world-sheet degrees of freedom \cite{delaOssa:1992vci}. Unlike its abelian counter part, the realization of the NATD as a symmetry of the full string theory is yet to be accomplished \cite{Alvarez:1993qi}-\cite{Sfetsos:1996pm}. The inclusion of the RR fluxes into the picture \cite{Sfetsos:2010uq} and subsequently establishing the connection with the Maldacena-Nunez conjecture \cite{Maldacena:2000mw} have provided enough motivation to study these backgrounds further.} (NATD) of $ AdS_5 \times S^5 $ \cite{Lozano:2017ole} is of special interest. This is due to the fact that NATD solution does not naturally lead to the so called ``background completion" and thereby corresponds to \emph{smeared} D0 branes near the asymptotic infinity which translates into an \emph{irrelevant} deformation on the dual field theory counterpart \cite{Lozano:2017ole}. These geometries, to start with, are therefore quite unusual in the sense that the vacuum of the dual QFT is not naturally the usual PWMM vacuum \cite{Berenstein:2002jq} that corresponds to relevant deformation of \cite{Banks:1996vh}. However, as authors show in \cite{Lozano:2017ole}, this problem can be cured and the background completion can be achieved following a ``coarse-graining" approximation \cite{Bak:2005ef}-\cite{Donos:2010va} of LM.

The NATD background mentioned above can be obtained by applying a non-Abelian T duality inside the $ AdS_5 $ of the full $ 10d $ spacetime. In the dual supergravity description, this vacuum is realized through the application of the NATD for the subgroup $ SU(2)(\subset SO(2,4)) $ inside the $ AdS_5 $. As far as the dual BMN Matrix model is concerned, these transformations correspond to preserving half of the original 32 supercharges (of $ \mathcal{N}=4 $ SYM on $ R\times S^3 $) by integrating out all the fields that transform under the given subgroup. 

 When uplifted to $ 11d $, this precisely corresponds to the Penrose limit of the superstar solution in $ AdS_7 \times S^4 $ \cite{Leblond:2001gn}. Following the methodology as developed in \cite{Polchinski:2000uf}, this Type IIA solution must satisfy a particular set of electrostatic boundary value problems \cite{Lin:2004kw}-\cite{Lin:2005nh} for this to be fit within the formalism of \cite{Lin:2004nb}.

The organization for the rest of the paper is as follows: In Section 2, we provide a detailed discussion on matrix models and their connection with Type IIA solutions in string theory. In Section 3, we calculate the Euclidean Wilson loops in BMN matrix model using the localization techniques. In Section 4, we discuss the stringy counterpart of these Wilson operators as open string embedding in NATD of $ AdS_5 \times S^5 $. Finally, we draw our conclusion in Section 5.
%%%%%%%%%%%%%%%%%%%%%%%%%%%%%%%%
\section{Matrix models and Type IIA strings: A review}
We first elaborate on the general idea behind the duality which was initiated by authors in their seminal work \cite{Berenstein:2002jq}. The field theory side of this duality is obtained by modding out the $ \mathcal{N}=4 $ SYM defined on $ R \times S^3 $ by subgroups of $ \hat{SU}(2) $. In other words, one integrates out fields that transform under the given subgroups of $ \hat{SU}(2) $.

Depending on the choices of these subgroups namely (i) $ Z_k $, (ii) $ U(1) $ and (iii) $ \hat{SU}(2) $ one lands up into a class of dual field theories. These theories are classified as (i) $ \mathcal{N}=4 $ SYM on $ R \times S^3/Z_k $, (ii) $ \mathcal{N}=8 $ SYM on $ R \times S^2$ and (iii) the BMN Plane Wave Matrix Model respectively. It is the example (iii) that we are interested throughout this paper.

The BMN Plane Wave Matrix Model enjoys a global $ SU(2|4) $ symmetry as well as preserves $ 16 $ supercharges. The different vacuum for this theory are given by fuzzy spheres that are classified into different representations of the $ SU(2) $ Lie algebra and are in a one to one correspondence with the partitions of $ N $ where $ N $ stands for the dimension of the $ SU(2) $ Lie algebra. In the dual Type IIA gravitational description, these partitions are realized in terms of the spherical NS5 branes which we elaborate now.

For each of these above vacua, the dual geometry was first proposed by authors in \cite{Lin:2005nh}. These are the class of geometries that enjoy a global $ R \times SO(3) \times SO(6) $ isometry. Remarkably enough, the supergravity equations governing these geometries could be recast into the form of Laplace equations of electrostatics for some axially symmetric electrostatic problem. The geometry is completely specified by a set of conducting disks (each of which corresponds to a specific vacuum of the BMN Plane Wave Matrix Model (PWMM) that are mentioned above) and a potential function ($ V(\sigma , \eta) $) satisfying Laplace's equation. Let us elaborate more on these ideas here. Typically, the Type IIA supergravity background is expressed as \cite{Lin:2005nh}
\begin{align}
\label{1}
&ds_{10}^2 = \left( \frac{\ddot{V}-2 \dot{V}}{V''}\right)^{1/2} \Big[ -\frac{4 \ddot{V}}{\ddot{V}-2 \dot{V}}dt^2 -\frac{2 V''}{\dot{V}}(d\sigma^2 \nonumber\\&+ d\eta^2)+4 d\Omega^{2}_5 +\frac{2 \dot{V} V''}{((\ddot{V}-2\dot{V})V''  -(\dot{V}')^2)} d\Omega^{2}_2\Big],
\end{align} 
where we define various derivatives as, $ \dot{V}=\sigma \partial_{\sigma}V $, $ V' =\partial_{\eta}V $, $ \ddot{V}=\sigma \partial_{\sigma}\dot{V} $ and $ V''=\partial^2_{\eta}V $.

The Type IIA background (\ref{1}), preserving $\mathcal{N}=2$ supersymmetries, is completely specified by the potential function $ V(\sigma , \eta) $ which solves the Laplace equation \cite{Lin:2004nb}, \cite{Lin:2005nh}
\begin{eqnarray}
\ddot{V}+\sigma^2 V'' =0.
\end{eqnarray}

As pointed out by authors in \cite{Lin:2005nh}, the PWMM corresponds to D0 branes plus some relevant deformations. In what follows, for the PWMM to be described at the boundary, the corresponding potential function takes the following form 
\begin{eqnarray}
\label{e2.9}
V_{PWMM}(\sigma , \eta)=  \eta \sigma^2 - \frac{2}{3}\eta^3 +\frac{P \eta}{(\eta^2 + \sigma^2)^{3/2}}=V_{back}+V_{dipole},
\end{eqnarray}
where $ P (\sim N_0) $ is the dipole moment produced by the conductiong disk which also counts the number of D0 branes \cite{Lin:2005nh}. 

Here, $ V_{back}\sim  \eta \sigma^2 - \frac{2}{3}\eta^3$ corresponds to the background electrostatic potential. On the other hand, $ V_{dipole} \sim \frac{P \eta}{(\eta^2 + \sigma^2)^{3/2}} $ is the leading diploe moment contributions (in the asymptotic limits) due to conducting disks. Clearly, given the potential (\ref{e2.9}), the asymptotic $ (\sigma , \eta)\rightarrow \infty $ limits of the metric (\ref{1}) behaves like that of a D0 brane. 

Here, $\eta= \eta_i =\frac{\pi}{2} N^{(i)}_5 $ corresponds to the location of a set of conducting disks with radii $R_i $ and charge $ Q_i \sim \frac{\pi^2}{8}N^{(i)}_2 $ \cite{Lin:2005nh}. In other words, the location of these disks are fixed by the NS5 branes in the bulk at strong coupling\footnote{\label{f3} From the perspective of the dual PWMM, one can consider a partition of the $ N_0 \sim N $ dimensional representation of the $ SU(2) $ Lie algebra as, $ N=\sum_i k_i N_2^{(i)}$ where $ k_i $s are the ranks of the smaller \emph{irreducible} representations of the full $ N $- dimensional representation. Here, $ N_2^{(i)} $ corresponds to the number of copies in each of these irreducible representation. Each of these partitions corresponds to a fuzzy sphere (with radii $ k_i $) that stands for one possible vacua of the BMN matrix model. In the dual super-gravity description, the partition of $ N $ into smaller representations stands for a polarization \cite{Myers:1999ps} of D0 branes into D2 branes (at weak coupling) or NS5 branes (at strong coupling) in the presence of background fluxes. At strong coupling, one can imagine a partition in the super-gravity description as $ N_0 \sim P \sim \sum_i N_{5}^{(i)} Q_i $. Here, $ N_{5}^{(i)}$ corresponds to dimension of a irreducible representation and $ Q_i \sim N_{2}^{(i)}$ corresponds to number of copies in each of these representations. In the present analysis, we consider a limit in which $ N_0 \sim P \gg 1 $ which therefore corresponds to irreducible representations with larger ranks. }. These disks are placed in a background electric field that grows to infinity in the asymptotic limits and vanishes as $ \eta \sim 0 $ which therefore corresponds to an infinite conducting plane. In the language of the the dual PWMM, this corresponds to a double scaling limit in which the t'Hooft coupling grows large and the corresponding rank of the gauge group goes to infinity.

Arguably recently, the authors in \cite{Lozano:2017ole} constructed a new class of $ \mathcal{N}=2 $ Type IIA vacua\footnote{The associated potential function and the background solutions are discussed in detail in Section \ref{sec4}.} by means of applying a non-Abelian T duality (NATD) to the $ AdS_5 $ subspace of the full $ AdS_5 \times S^5 $ background that fall under the category of \cite{Lin:2005nh}. Nonetheless, in their construction \cite{Lozano:2017ole}, the authors provide a proper completion of the NATD background in terms of smeared D0 branes and interpret these solutions in terms of a $ (0+1) $ dimensional QFT which is conjectured to be a relevant deformation of the BMN Plane Wave Matrix Model \cite{Berenstein:2002jq}. The purpose of the present paper, is to probe through this duality conjecture by estimating observable on both sides of the duality and compare the findings thereafter.  
%%%%%%%%%%%%%%%%%%%%%%%%%%%%%%%
\section{Wilson loops from matrix models}
We begin our analysis with a model calculation for the Euclidean Wilson loops in BMN matrix model using the localization techniques in which the eigen value distributions of the adjoint scalar is identified with the charge distribution over the conducting discs in the dual gravity description. 
Wilson loops in PWMM were first proposed using an $ S^3 $ reduction by authors in \cite{Ishiki:2011ct} which were further investigated in \cite{Asano:2012zt}. We begin by reviewing the essentials of PWMM \cite{Berenstein:2002jq} in $ (0+1)d $ which is described by the following action \cite{Asano:2012zt}-\cite{Asano:2014vba}
\begin{eqnarray}
\label{4}
S= \frac{1}{g^2}\int d\tau ~ \text{Tr} (\frac{1}{4}F_{MN}F^{MN}+\frac{m^2}{8}X_m X^m +\frac{i}{2}\Psi \Gamma^M D_M \Psi),
\end{eqnarray}
where we label the indices as, $ M=\lbrace1, ~a,~m \rbrace$ where $ a=2,3,4 $ and $ m=5, \cdots, 10 $. Here, $ X_1 $ is referred as the one dimensional gauge field together with $ X_a $ and $ X_m $ as $ SO(3) $ and $ SO(6) $ scalars respectively. Moreover, the entities $ F_{MN} $ and $ D_M \Psi $ are defined as follows
\begin{align}
&F_{1M}= \partial_1 X_M -i [X_1 , X_M]~;~F_{ab}=m \epsilon_{abc}X_c -i [X_a , X_b]\\
&F_{am}=-i [X_a , X_m]~;~F_{mn}=-i[X_m , X_n]\\
&D_1 \Psi = \partial_1 \Psi - i[X_1 ,\Psi]~;~D_m \Psi = - i[X_m ,\Psi]\\
&D_a \Psi = \frac{m}{8}\epsilon_{abc}\Gamma^{bc}\Psi -i [X_a , \Psi ],
\end{align}
where we define the derivative $ \partial_1 $ with respect to $ \tau $.

The model (\ref{4}) preserves $ 16 $ supersymmetries. On top of that, it possesses $ SO(6) $ global symmetries corresponding to the rotation among the $ X_m $ scalars and a global $ SO(3) $ symmetry that rotates $ X_a $ scalars. Combining these together, the model (\ref{4}) enjoys a global $ R \times SO(3) \times SO(6) $ symmetry as a part of the bosonic subgroup of $ SU(2|4) $. Notice that, this is precisely the symmetry as realized on the dual gravitational counterpart.

The vacuum of the model (\ref{4}) preserves $ SU(2|4) $ symmetry and are given by fuzzy spheres that were mentioned earlier. The classical vacuum of the model are labeled by $ SU(2) $ representations and are obtained by setting all others except $ X_a $ to zero
\begin{eqnarray}
\label{e5.2}
X_a = -2L_a =-2 \bigoplus_{s=1}^{\Lambda }1_{N_2^{(s)}}\otimes L_a^{[N_5^{(s)}]},
\end{eqnarray}
where $ \Lambda  $, $ N_2^{(s)} $ and $ N_5^{(s)} $ are integers. On the other hand, $ L_a^{[N]}$ is the generator in the $ N $ dimensional irreducible representation of $ SU(2) $ which satisfies the corresponding Lie algebra $ [L_a^{[N]},L_b^{[N]}] =i\epsilon_{abc}L_c^{[N]}$ in its standard form.

The classical vacuum (fuzzy spheres) of the theory are realized as a partition of $ N \sim \sum_{s=1}^{\Lambda }N_2^{(s)} N_5^{(s)}$ where $ N_2^{(s)} $ stands for the multiplicity for the $ N_5^{(s)} $ dimensional irreducible representation of $ SU(2) $. In what follows, we work in a limit in which these multiplicities are taken to be large enough so that instanton contributions can be ignored. 

The purpose of the present exercise is to compute VEV for Wilson operators using Euclidean partition function. These Wilson operators, as we show in the next Section, are precisely the Euclidean Wilson loops constructed using open string embedding in NATD of $ AdS_5 \times S^5 $. The computation on the matrix model side requires to set the boundary conditions at $ \tau \rightarrow \pm \infty $ such that all the fields reach vacuum configurations in these asymptotic limits. In the limit when the multiplicities ($ N_2^{(s)}\gg 1 $) are large enough, the instanton effects can be ignored and as a result, the Euclidean partition function is well defined around these fixed vacuum near the asymptotic infinities.

Under these assumptions one can compute the coorelators made out of $ \phi (\tau) $ using the techniques of localisation\footnote{This comes along with the assumption that the fields reach classical vacuum at $ \tau \rightarrow \pm \infty $.} \cite{Asano:2012zt}-\cite{Asano:2014vba}.  Here $ \phi (\tau) $ is a complex scalar field which is a linear combination of both $ SO(3) $ as well as $ SO(6) $ scalars. For example, one can choose \cite{Asano:2014vba}
\begin{align}
\phi (\tau)=2(-X_4 (\tau)+\sinh \tau X_9 (\tau)+i \cosh \tau X_{10}(\tau))
\end{align}
where $ X_4 (\tau) $ is a $ SO(3) $ scalar while the remaining are the $ SO(6) $ scalars.

Using the result of localization, one can estimate a $ n $ point correlation function for a given a smooth function $ f(\phi (\tau)) $ which reveals the following equality
\begin{eqnarray}
\label{e5.5}
\langle \prod_{a'=1}^{n} \text{Tr} f_{a'}(\phi (\tau_{a'}))\rangle_{L_a} = \langle \prod_{a'=1}^{n} \text{Tr} f_{a'}(L_a +iM) \rangle_{mm},
\end{eqnarray}
were the L.H.S. is estimated around the vacuum (\ref{e5.2}) of PWMM at infinity. On the other hand, here $ M :=\bigoplus_{s=1}^{\Lambda }1_{N_2^{(s)}}\otimes M_{s}$ where $ M_s $ is an $ N_2^{(s)} \times N_2^{(s)} $ hermitian matrix that commutes with all three generators $ L_a^{[N_{5}^{(s)}]} $ ($ a=1,2,3 $) of $ SU(2) $. 

The correlator on the R.H.S. of (\ref{e5.5}) could estimated with reference to the partition function that serves as the starting point of our analysis
\begin{eqnarray}
\label{e5.6}
\mathcal{Z}:=\int \prod_{j=1}^{N_2}dm_j \prod_{j>i}^{N_2}\Xi (m_i , m_j)e^{-\frac{2N_5}{g^2}\sum_{j=1}^{N_2}m^2_j},
\end{eqnarray}
where we choose to work with the $ \Lambda =1 $ case which corresponds to a one matrix model as discussed in \cite{Lin:2005nh}. Here, $ m_j $s are the eigen values of the matrix $ M $.

The function in (\ref{e5.6}) is defined as \cite{Asano:2014vba}
\begin{eqnarray}
\Xi (m_i , m_j)=\prod_{J=0}^{N_5 -1}\frac{((2J +2)^2+(m_j -m_i)^2)(4J^2 +(m_j -m_i)^2)}{((2J +1)^2+(m_j -m_i)^2)^2}.
\end{eqnarray}

The partition function (\ref{e5.6}) could be recast as
\begin{eqnarray}
\mathcal{Z}:=\int \prod_{j=1}^{N_2}dm_j e^{-\mathcal{S}_{E}(m)},
\end{eqnarray}
where the Euclidean action could be formally expressed as
\begin{eqnarray}
\label{e5.9}
\mathcal{S}_{E}(m):=\frac{2N_5}{g^2}\sum_{j=1}^{N_2}m^2_j -\sum_{j\neq i} \log \Xi (m_i , m_j).
\end{eqnarray}

The first term in the effective action (\ref{e5.9}) corresponds to the potential energy $ V(m_j) $ for $ N $ eigen values ($ m_1,\cdots,m_N $). The interaction between these eigen values corresponds to the logarithmic term in (\ref{e5.9}). 

It is interesting to notice that the interaction is strongly repulsive as these eigen values approach each other \cite{Dijkgraaf:2002fc}. This is consistent with their fermionic behaviour \cite{Asano:2014vba} and is consequence of the Pauli exclusion principle. In fact, a closer analysis reveals that 
\begin{eqnarray}
\log \Xi (m_i , m_j)|_{m_i \sim m_j}\sim  \log(\pi ^2  \Gamma (N_5)^2 (m_i -m_j)^2),
\end{eqnarray}
which indicates that the Coulomb repulsion between these eigenvalues increases logarithmically as they approach each other.

We now look for saddle points for (\ref{e5.9}). Considering large $ N_5 (\rightarrow \infty) $ limit we obtain
\begin{eqnarray}
\label{e5.11}
\frac{2N_5}{g^2}V'(m_i) -2 \pi\sum_{j \neq i}  \text{csch}(\pi  (m_i-m_j))\approx 0.
\end{eqnarray}

To proceed further, we further introduce the resolvent \cite{Dijkgraaf:2002fc}
\begin{eqnarray}
w(z)=\text{Tr}\left( \frac{1}{M-z}\right) =\frac{1}{N_5}\sum_{j=1}^{N_2}\frac{1}{m_j -z}.
\end{eqnarray}

Considering $ |m_i -m_j|<1 $ we finally obtain a differential equation for the resolvent
\begin{eqnarray}
\label{e5.13}
\frac{4N}{g^2}+\frac{4N^2_5 z}{g^2}w(z)+N_5^2w^2(z) -N_5 w'(z)\approx 0,
\end{eqnarray}
where $ N= N_5 N_2 $ corresponds to the size of the matrices in the matrix model.

Let us now explore the solution of (\ref{e5.13}) in the large $ N $ limit. Clearly, the first two terms on the L.H.S. of (\ref{e5.13}) are of $ \mathcal{O}(N) $. On the other hand, a closer look further reveals that $N_5^2 w^2 \sim \mathcal{O}(N^2_2) $ and $ N_5 w' \sim  \mathcal{O}(N_2)$. Therefore, considering the fact $ N \gg N_2 $ one can simply ignore the last term in (\ref{e5.13}) which finally yields a solution of the form
\begin{eqnarray}
\label{e5.14}
w(z)\sim \frac{2}{N_5^2} \left(-\frac{N_5^2 z}{g^2} \pm \frac{N_5\sqrt{N_5^2 z^2-g^2 N}}{g^2}\right) .
\end{eqnarray}

Using (\ref{e5.14}), the density of eigenvalues \cite{Dijkgraaf:2002fc} finally turns out to be
\begin{eqnarray}
\varrho (x)=\frac{2}{\pi g^2} \left(1-\frac{N_5 x}{\sqrt{g^2 N -N_5^2 x^2}}\right).
\end{eqnarray}

Clearly, (\ref{e5.11}) corresponds to $ N_2 $ equations which therefore has $ N_2 $ classical eigen values $ m_j(0) (j=1,\cdots,N_2) $ as solutions such that $ m_1(0)>m_2(0)>\cdots>m_{N_2}(0) $. Let us consider that $ m_1(0)=g $ be the classical eigen value for the first eigen mode and $ m_k (0)= g \cos\theta_k $ is the eigen value for the $ k $ th mode.

Total number of eigen modes between $ m_1(0) $ and $ m_k(0) $ is therefore given by
\begin{eqnarray}
k \sim \frac{2(1- \cos \theta_k )}{\pi  g}+\mathcal{O}(g^{-2}).
\end{eqnarray}

The VEV of Wilson loop operator in PWMM can be estimated through matrix integral using localisation techniques. At the saddle point, the Wilson operator (obtained through dimensional reduction) in PWMM turns out to be \cite{Asano:2012zt}
\begin{eqnarray}
\mathcal{W}_{PWMM}:=\frac{N_5}{N}\sum_{j=1}^{N_2}e^{2\pi m_j},
\end{eqnarray}
where we restrict ourselves to the one matrix model as mentioned earlier.

The VEV is therefore obtained through the matrix integral as
\begin{eqnarray}
\label{e5.18}
\langle \mathcal{W}\rangle_{mm}:=\frac{N_5}{N}\frac{1}{\mathcal{Z}}\int \prod_{j=1}^{N_2}dm_j \sum_{j=1}^{N_2}e^{2\pi m_j} e^{-\mathcal{S}_{E}(m)}.
\end{eqnarray}

The above expression (\ref{e5.18}) one could think of a sum of expectation values of the form $ \langle e^{2\pi m_i}\rangle_{mm} $ in the matrix model. With an assumption that all these expectation values are equivalent to each other, the above expression eventually simplifies to $ N_2  \langle e^{2\pi m_k}\rangle_{mm}  $. As a result, the corresponding VEV turns out to be
\begin{eqnarray}
\langle \mathcal{W}\rangle_{mm}:=\frac{1}{\mathcal{Z}}\int \prod_{j=1}^{N_2}dm_j e^{-\mathcal{S}(m)}.
\end{eqnarray}
where we define an effective action in the presence of source as
\begin{eqnarray}
\label{e5.20}
\mathcal{S}(k, m):=\frac{2N_5}{g^2}\sum_{j=1}^{N_2}m^2_j -\sum_{j\neq i} \log \Xi (m_i , m_j)-2\pi m_k.
\end{eqnarray}

We now want to explore the classical saddle point(s) of (\ref{e5.20}). Clearly, the classical equation of motion is given by
\begin{eqnarray}
\label{e5.21}
\frac{\delta \mathcal{S}}{\delta m_i}=0=\frac{4N_5}{g^2}m_i -2 \pi\sum_{j \neq i}  \text{csch}(\pi  (m_i-m_j))-2\pi \delta_{ik},
\end{eqnarray}
whose solutions we denote as $ m_j(k) (j=1,\cdots,N_2) $.

Assuming $ k $ as a continuous variable and the difference $ | \frac{m_j(k)-m_j(0)}{m_j(0)}|\ll 1 $, we find\footnote{Notice that, this is a valid assumption as the last term in (\ref{e5.21}) can be ignored in the limit $ N \gg N_2 $.}
\begin{eqnarray}
\frac{\partial \mathcal{S}(k, m)}{\partial k}=-m_{k}(0)=-g \cos \theta_{k}.
\end{eqnarray}

Applying the simple chain rule and imposing the boundary condition \cite{Yamaguchi:2006tq} $ \mathcal{S}(k=0)=0 $
\begin{eqnarray}
\frac{\partial \mathcal{S}(k, m)}{\partial \theta_k}|_{k=N}=\frac{\partial k}{\partial \theta}\frac{\partial \mathcal{S}(k, m)}{\partial k}|_{k=N},
\end{eqnarray}
one can finally show that for Euclidean Wilson loops of rank $ N $ 
\begin{eqnarray}
\mathcal{S}(k \sim N)\sim \frac{1}{g}\sin^2\theta_N.
\end{eqnarray}

As we show in the next Section, the $ \sin^2\theta$ behavior can be reproduced using open string solutions in NATD of $ AdS_5 \times S^5 $. 
%%%%%%%%%%%%%%%%%%%%%%%%%%%%
\section{Open strings in NATD of $ AdS_5 \times S^5 $}
\label{sec4}
The NATD of $ AdS_5 $ \cite{Lozano:2017ole} is characterised by the potential function
\begin{eqnarray}
\label{26}
V(\sigma , \eta)=-\eta \log\sigma +\frac{\eta \sigma^2}{2}-\frac{\eta^3}{3}.
\end{eqnarray}

It is due to the presence of the logarithmic term, the potential (\ref{26}) does not approach D0 brane solution in the asymptotic ($ \eta =const. , \sigma \rightarrow \infty $) limit of coordinates. Rather, it corresponds to D0 branes \emph{smeared} over $ \mathbb{R}^3 $. In the language of the BMN matrix model, this implies turning on \emph{irrelevant} deformation in the dual field theory \cite{Lozano:2017ole}. 

Based on the ``coarse-graining" approximations of LM \cite{Lin:2005nh}, the authors cure the above problem and propose a completion of the NATD background by introducing the potential function\footnote{The NATD background (\ref{26}) can be achieved in the strict $ L=\infty $ limit of the generalised potential function (\ref{e3.46}). In other words, the NATD background (\ref{26}) is a special (large $ L $) limit of (\ref{e3.46}) which produces smeared D0 branes in the asymptotic limit of the $ \sigma $ coordinate.} of the following form \cite{Lozano:2017ole}
\begin{eqnarray}
\label{e3.46}
V(\sigma , \eta )=(\eta \sigma^2 -\frac{2}{3}\eta^3)-2 \alpha  \eta  \log \sigma + \alpha  \sqrt{(L-\eta )^2+\sigma ^2}-\alpha  \sqrt{(\eta +L)^2+\sigma ^2}\nonumber\\
+\alpha  \eta  \log \left(\left(\sqrt{(L-\eta )^2+\sigma ^2}+(L-\eta )\right) \left(\sqrt{(\eta +L)^2+\sigma ^2}+(\eta +L)\right)\right),
\end{eqnarray}
where the coordinate $ \eta \in [0, L] $ lives in a finite interval to start with. 

Given the potential function as in (\ref{e3.46}), the corresponding metric and the NS-NS two form can be expressed as \cite{Lozano:2017ole}
\begin{eqnarray}
\label{e2.2}
ds^2_{10}&=&-f_1(\sigma , \eta) dt^2 +f_2 (\sigma , \eta) (d\sigma^2 + d\eta^2)+f_3 (\sigma , \eta) d\Omega^2_2 (\theta , \varphi)+f_4 (\sigma , \eta)  d\Omega^2_5 ,\\
\mathcal{B}_2 &=& f_6(\sigma , \eta)d\Omega_2 (\theta , \varphi),
\end{eqnarray}
where the individual coefficients $ f_i (\sigma , \eta) $ can be expressed as
\begin{eqnarray}
f_1(\sigma , \eta)&=&\frac{4 \ddot{V}}{\sqrt{V''(2\dot{V}-\ddot{V})}}~;~f_2(\sigma , \eta)=\frac{8 \ddot{V}}{\dot{V}}\frac{1}{f_1},\\
f_3(\sigma , \eta)&=&-\frac{8 \dot{V}\ddot{V}}{\Delta}\frac{1}{f_1}~;~f_4(\sigma , \eta)=-\frac{16 \ddot{V}}{V''}\frac{1}{f_1},\\
f_5(\sigma , \eta)&=&\frac{4(2\dot{V}-\ddot{V})^3}{V'' \dot{V}^2 \Delta^2}~;~f_6(\sigma , \eta)=2\left( \eta + \frac{\dot{V}\dot{V}'}{\Delta}\right),\\
f_7(\sigma , \eta)&=& \frac{2 \dot{V}\dot{V}'}{2\dot{V}-\ddot{V}}~;~f_8(\sigma , \eta ) = -\frac{4\dot{V}^2 V''}{\Delta},
\end{eqnarray}
together with, $ \Delta =(\ddot{V}-2\dot{V})V''  -(\dot{V}')^2$.

In order to compute the expectation value of the circular Wilson loop at the boundary, one has to compute the minimal area of the string world-sheet that ends on this circle. For our calculation this circle is the \emph{Euclidean} time circle $\tau $. 

Let us choose to work with open string embedding whose ends are lying on $ S^2 $ and which is extended inside the bulk along $ \sigma $. In what follows, we choose to work with the Polyakov action of the sigma model
\begin{eqnarray}
\label{e2.10}
S_{P}&=&\frac{\sqrt{\lambda}}{2\pi}\int d^2 \sigma \mathcal{L}_{P},\\
 \mathcal{L}_{P}&=&\eta^{ab}G_{MN}\partial_{a}X^M\partial_{b}X^N+\mathcal{B}_{MN}\partial_{a}X^M\partial_{b}X^N \varepsilon^{ab}.
\end{eqnarray}

We now keep $\sigma  = \sigma_c$ fixed and estimate Wilson loops for the full $ N $ dimensional representation of $ SU(2) $ where the density of the irreducible representations (in the partition of $ N $) are bounded by the interval $ \eta \in [0,L] $.

Considering an embedding for which $\sigma = \sigma_c \gg 1 $
\begin{eqnarray}
t =0 ~;~ \sigma = \sigma (\sigma^1)~;~\eta = \eta (\sigma^1) ~;~\theta = \theta (\sigma^1 )~;~\varphi =\sigma^0 = \tau ,
\end{eqnarray}
the resulting metric functions and NS-NS two form become
\begin{eqnarray}
f_1 (\sigma = \sigma_c , \eta) &=& \frac{4 \sigma^{\frac{7}{2}}_c}{L^{\frac{3}{2}}}\sqrt{\frac{2}{5}}~;~f_2 (\sigma \sim \sigma_c , \eta)  = \frac{2 \sqrt{10}L^{\frac{3}{2}}}{ \sigma^{\frac{7}{2}}_c},\\
f_3 (\sigma \sim \sigma_c , \eta) & = & \frac{2 \sqrt{10}L^{\frac{3}{2}}\eta^2}{ \sigma^{\frac{7}{2}}_c}~;~f_6 (\sigma \sim \sigma_c , \eta) = \frac{30 \eta ^3 L^3}{\sigma_c ^7}.
\end{eqnarray}

Virasoro constraints, on the other hand, are given by
\begin{eqnarray}
\label{e2.31}
T_{\sigma^1 \sigma^1} &=&- f_3 \sin^2\theta +f_2 (\eta'^2 + \sigma'^2 ) +f_3 \theta'^2 =0,\\
T_{\sigma^0 \sigma^1} &=&0.
\end{eqnarray}

Using Virasoro constraint (\ref{e2.31}), this further yields
\begin{eqnarray}
\label{41}
\theta' = \pm \sqrt{|\sin^2\theta -1|}~;~\eta' =-\sqrt{\frac{f_3}{f_2}}.
\end{eqnarray}

After a straightforward substitution we find
\begin{eqnarray}
S^{(reg)}_P = \frac{4\sqrt{10}L^{\frac{3}{2}}\sin^2\theta_{0}\sqrt{\lambda}}{\sigma^{\frac{7}{2}}_c}\int_{0}^{L}d\eta ~\eta =  \frac{2\sqrt{10}L^{\frac{7}{2}}\sin^2\theta_{0}\sqrt{\lambda}}{\sigma^{\frac{7}{2}}_c},
\end{eqnarray}
where we fix the angle $ \theta = \theta_0 (=\frac{\pi}{2})$ which is a solution of (\ref{41}). Clearly, this solution corresponds to open strings that are extended from the north pole of $ S^2 $ to the equator whose end points define a loop along the equatorial plane.

This yields the corresponding Wilson operator of the form
\begin{eqnarray}
\langle \mathcal{W}_E \rangle = e^{-\sqrt{\tilde{\lambda}}\sin^2\theta_{0}}=e^{-S^{(reg)}_P},
\end{eqnarray}
where we introduce the rescaled coupling of the form $ \sqrt{\tilde{\lambda}}=2\sqrt{10}\left(\frac{L}{\sigma_c} \right)^{\frac{7}{2}} \sqrt{\lambda}$.
%%%%%%%%%%%%%%%%%%%%%%%%%%%%%%%%%%
\section{Concluding remarks}
The present paper is an attempt to test the celebrated ``holographic framework" in the light of estimating various non-local observable (for example, Wilson loops) in Large N gauge theories and comparing them with string theory calculations. One such example that has been considered in this paper is the conjectured duality between the BMN Plane Wave Matrix Model (PWMM) and the string theory description in Type IIA supergravity preserving $\mathcal{N}=2$ super-symmetry. 

We show that there is a particular class of Euclidean BMN Wilson operators of the PWMM that allow a dual description in terms of the open string solutions in non-Abelian T dual (NATD) of $ AdS_5 \times S^5 $. We calculate these Wilson operators using both descriptions and show that they allow a universal relation of the form $ \langle \mathcal{W}_E \rangle \sim e^{-\kappa \sin^2\theta} $ where $ \kappa $ stands for the coupling constants of the respective theories. These results therefore serve as the direct test for the duality conjecture in some Type IIA set-up and certainly validate the applicability of the holographic correspondence. 

A careful look reveals that the Wilson loop is extended along the unbroken two sphere $S^2(\theta , \varphi)$ of $AdS_5 \times S^5$. In other words, the NATD does not touch the above two sphere of the parent $AdS_5 \times S^5$ background. It is therefore possible to construct a dual Wilson loop by extending an open string along this unbroken $S^2(\theta , \varphi)$.
%%%%%%%%%%%%%%%%%%%%%%%%%%%%%%%%%%%%%%%%%%%%%%%%%%%%%%%%%%%%%%%%%%%%%%
\paragraph{Acknowledgements.}  I am indebted to the authorities of IIT Roorkee for their unconditional support towards researches in basic sciences. I acknowledge The Royal Society, UK for financial assistance. I also acknowledge the Grant (No. SRG/2020/000088) received from The Science and Engineering Research Board (SERB), India. 
%%%%%%%%%%%%%%%%%%%%%%%%%%%%%%%%%%%%

\end{document}